\documentclass[aps,floatfix,nofootinbib]{revtex4}

\usepackage{graphicx}
\usepackage{epic}
\usepackage{eepic}
\usepackage{latexsym}

\newcommand{\eq}[1]{(\ref{#1})}
\newcommand{\be}{\begin{equation}}
\newcommand{\ee}{\end{equation}}
\newcommand{\bea}{\begin{eqnarray}}
\newcommand{\eea}{\end{eqnarray}}

\newcommand{\vs}[1]{\vspace{#1 mm}}
\newcommand{\hs}[1]{\hspace{#1 mm}}

\newcommand{\bk}{{\bf k}}
\newcommand{\bx}{{\bf x}}
\newcommand{\bkp}{{\bf k'}}

\def\d{\delta}

\def\fr{\frac}

\def\m{\mu}
\def\n{\nu}

\def\r{\rho}
\def\s{\sigma}

\def\o{\omega}

\def\del{\partial}

\let\bm=\bibitem
\def\nn{\nonumber}

\begin{document}
\large

\title{ \Large Uncertainty relations for cosmological particle creation and \\
existence of large fluctuations in reheating}

\author{Ali Kaya}
\email[]{ali.kaya@boun.edu.tr}
\affiliation{\large Bo\~{g}azi\c{c}i University, Department of Physics, \\ 34342,
Bebek, Istanbul, Turkey }

\begin{abstract}
\large

We derive an uncertainty relation for  the energy density and pressure of a quantum scalar field in a time-dependent, homogeneous and isotropic, classical  background, which implies the existence of large fluctuations comparable to their vacuum expectation values. A similar uncertainty relation is known to hold  for the field square since the field can be viewed as a Gaussian random variable. We discuss possible implications of these results for the reheating process in  scalar field driven inflationary models, where reheating is achieved by the decay of the coherently oscillating inflaton field. Specifically we argue that the evolution after backreaction can seriously be altered by the existence of these fluctuations. For example, in one model the coherence of the inflaton oscillations is found to be completely lost in a very short time after backreaction starts. Therefore we argue that 
entering a smooth phase in thermal equilibrium is questionable in such models and reheating might destroy the smoothness attained by inflation.  
 
\end{abstract}

\maketitle
\vs{40}

\centerline{Essay written for the Gravity Research Foundation 2011 Awards for Essays on Gravitation}

\vs{20}

\centerline{Received Honorable Mention}


\newpage

Our universe is very special. The smoothness of the cosmic microwave background
(CMB) radiation is the prominent evidence for this assertion. It is remarkable
that CMB photons  were in thermal equilibrium on scales much larger than the
horizon size at the time of decoupling. Therefore,  causality precludes
thermalization of the CMB radiation by local interactions and the problem we are
facing is completely different than a box of gas reaching thermal equilibrium.
Actually,  there is a deeper difficulty since in the standard cosmological model
without inflation, most of the  CMB photons did not have a chance to causally
communicate at all.  

Inflation claims to solve this problem by exponentially enlarging a small causal
patch, which later  becomes  our observed universe. However, when inflation ends
one does not observe thermal equilibrium right away. Rather, in the scalar field
driven models for example, one finds an extremely smooth, flat universe filled
only with a coherently oscillating inflaton field; all rest that existed before
inflation are red-shifted away by the exponential expansion. Thermal
equilibrium must be reached  by the interactions accompanying the decay of the
inflaton field.    

This shows that one still needs to understand  the thermalization process
in an expanding universe, which is a very non-trivial and difficult problem to
study (see \cite{ber}). To have a correct picture in this setup, it is very crucial to determine the state of the universe just before the thermalization starts.  Our aim in this essay is to point out some important features of quantum particle creation in a classical background, which must be taken into account during reheating and before thermalization. 

The particle creation process during reheating  is usually studied using Fourier
decomposition and momentum modes (see e.g. \cite{str}). While it is perfectly legitimate to use
momentum modes, or in essence any complete set of modes, one must be careful
about a few potential issues. Firstly, causality requires the particle creation
process to take place independently in each horizon. Thus, in physically
interpreting the creation of a globally defined momentum mode one must ensure
that locality is not broken down by hand.  Secondly, there is a subtlety in the
definition of the number density of a completely dislocalized momentum mode.
Finally, to make sense of local physical quantities  a suitable
regularization must be utilized and this may also change some results obtained
in the momentum space. An important example of this kind is the modification of
the power spectrum by adiabatic regularization, studied recently in
\cite{p1}. All these issues can be bypassed if the particle creation
effects are described by giving the vacuum expectation value of the
energy-momentum tensor, which we adapt in the following. 

Consider now  a real  scalar field $\chi$ propagating in a Friedmann-Robertson-Walker  metric  
\be\label{m}
ds^2=-dt^2+a^2(dx^2+dy^2+dz^2),
\ee
which has the following action
\be\label{ac1}
S=-\fr12 \int \sqrt{-g}\left[(\nabla\chi)^2+M^2 \chi^2\right].
\ee
We assume that the mass parameter $M$ may also depend on time: $M=M(t)$, and therefore particle creation effects occur due to the time dependence of the scale factor $a$ and the externally varying  mass parameter $M$. For quantization, it is convenient to define
$X=a^{3/2}\chi$, and introduce the Fourier modes and time-independent  creation-annihilation
operators as  
\bea\label{exp}
X=\fr{1}{(2\pi)^{3/2}} \int d^3 k\left[ a_\bk X_k e^{-i\bk.\bx}+  a_\bk^\dagger
X_k^* e^{i\bk.\bx}\right],
\eea
where $[a_\bk,a^\dagger_\bkp]=\d(\bk-\bkp)$, $X_k$ obeys the Wronskian condition $X_k\dot{X}_k^*-X_k^*\dot{X}_k=i$ and $\ddot{X}_k+\o_k^2X_k=0$ with the frequency defined as  $\o_k^2=M^2+\fr{k^2}{a^2}-\fr94 H^2
-\fr32 \dot{H}$. 

The ground state of the system is defined as $a_\bk|0>=0$. 
It is a straightforward exercise to calculate the vacuum expectation value
of the energy momentum tensor $<T_{\m\n}>=<0|T_{\m\n}|0>$ corresponding to the 
$\chi$-particles created on the background. By noting that the energy density
and the pressure of the $\chi$-particles are given by   
 \bea
\r(t,{\bf x})&=&\fr{1}{2}\dot{\chi}^2+\fr{1}{2}g^{ij}(\del_i\chi)(\del_j\chi)+\fr{1}{2}
M^2\chi^2,\nn\\
P(t,{\bf x})&=&\fr{1}{2}\dot{\chi}^2-\fr{1}{6}g^{ij}(\del_i\chi)(\del_j\chi)-\fr{1}{2}
M^2\chi^2,\label{ex}
\eea
one can obtain 
\be\label{vev}
<\r>=T+V+G,\hs{5}
<P>=T-V-\fr{1}{3}G,\hs{5}
\ee
where $T$, $V$ and $G$ are given by
\bea
T&=&\fr{1}{2(2\pi a)^3}\int d^3 k\, |\dot{X}_k-\fr{3}{2}HX_k|^2,\nn\\
V&=&\fr{M^2}{2(2\pi a)^3}\int d^3 k\,\, |X_k|^2,\label{func}\\
G&=&\fr{1}{2(2\pi a)^3}\int d^3 k \,\fr{k^2}{a^2}\,|X_k|^2.\nn
\eea
Also, it is of some interest to determine the field
variance  $<\chi^2>=<0|\chi^2(t,{\bf x})|0>$, which can be found as
\be\label{ce}
<\chi^2>=\fr{1}{2(2\pi a)^3}\int d^3 k\, |X_k|^2.
\ee 
It is clear that  $T$ and $V$ can be interpreted as the kinetic and the potential
energies, respectively, and $G$ can be viewed as the energy stored in the
gradient of the $\chi$-field. In general all the above expectation values
diverge and thus a suitable regularization must be applied. After
regularization, however,  the functions $T$, $V$, $G$  and $<\chi^2>$ can be
made finite. Note that since the background is homogenous the vacuum expectation
values \eq{vev} and \eq{ce} do not depend on the spatial coordinates.  

The above expressions only give the average values of the corresponding physical
quantities and one would expect to determine (quantum) fluctuations about these
averages. In \cite{a3}, we explicitly calculate them. For example, after defining the fluctuation operators $\d \rho=\rho(t,{\bf x})-<\rho>$ and
$\d P=P(t,{\bf x})-<P>$, a relatively long but straightforward calculation, which uses 
\eq{exp} and \eq{ex},
 gives
\be\label{f1}
<\d\r^2>+<\d P^2>=4T^2+4V^2+\fr{20}{27}G^2.
\ee
From \eq{vev} this result shows that the deviations in the energy density and the pressure are {\it
always} of the order of the average energy density $<\rho>$. Similarly, although
their vacuum expectation values vanish one can see that the momentum density 
and the stress tensor have non-zero variances  proportional to $TG$ and $G^2$,
respectively \cite{a3}. Thus, if $T$ and $G$ are not small compared to $V$,  the
deviations in the momentum density and the stress can also be as large as
$<\rho>$.  Eq. \eq{f1} can be viewed as an uncertainty relation involving energy density and pressure, where the order of magnitude of the uncertainty is fixed by the average energy density. 

It is easy to see that there also exists order one fluctuations in  $\chi^2$, i.e. the
field variance has a large variance. Defining $\d\chi^2=\chi^2(t,{\bf x})-<\chi^2>$,
one can easily calculate 
\be\label{f2}
\sqrt{<(\d\chi^2)^2>}=\sqrt{2}<\chi^2>.
\ee
This result directly follows from the fact that $\chi(t,{\bf x})$ is a Gaussian
random variable defined at the point ${\bf x}$ and \eq{f2} is true for any
Gaussian distribution.  

The fluctuations in \eq{f1} and \eq{f2}, which are actually defined pointwise, 
should be interpreted as the statistical averages over different points in space
at a given time. However, it is crucial to realize  that  the field
variables at nearby points are {\it not statistically independent}. The (comoving) 
size of a typical region containing correlated field variables is given by the correlation
length $\xi_c$, which is the minimum value of $|{\bf x}|$ such that  the
two-point function  vanishes $<\chi(t,{\bf x})\chi(t,0)>=0$. The correlation
length is one of the most important parameters characterizing quantum
fluctuations. One can imagine the space to be divided into regions of typical
size $\xi_c$, and in each region the physical quantities $\rho$, $P$ and
$\chi^2$ can be thought to be  uniformly distributed. The variations given
in \eq{f1} and \eq{f2} characterize  changes from one such  region  to another.   

It is interesting to compare these findings with \cite{a1,a2} where we study the particle creation process using a complete orthonormal family of localized wave-packets and find out existence of fluctuations in the number and the energy densities of
particles produced in any given volume $V$.  As noted in  \cite{a1,a2},
if the volume $V$ is sufficiently large, the relative deviations of these
perturbations fall down like $1/\sqrt{V}$. This behavior can easily be
understood since the total deviation is due to  $V/v_0$ number of statistically
independent random variables, where $v_0$ is the volume occupied by each random
variable, which can be fixed in terms of the correlation length as $v_0=\xi_c^3$. 

One may now wonder if the presence of these fluctuations can have any cosmological significance. Here, we consider the reheating process in a single scalar field model. Assume that near its minimum the potential of the inflaton field takes the form 
$\fr{1}{2} m^2\phi^2$, where $m$ is the inflaton mass. In that case the
inflaton $\phi(t)$ and the scale factor of the universe $a(t)$ can be determined as  
\be\label{inf}
\phi=\Phi(t)\sin (m t), \hs{10} \Phi(t)\simeq
\fr{C}{t},\hs{10}a\simeq\left(\fr{t}{t_0}\right)^{2/3}.  
\ee
The pressureless dust equivalent expansion in \eq{inf} can be understood as follows: As the
inflaton oscillates about its minimum, the kinetic and the potential energy
terms consecutively dominate the energy-momentum tensor. While the kinetic
energy dominance can be described by the equation of state $P=\rho$ the
potential energy dominance implies $P=-\rho$. Thus on the average one gets
$P=0$. 

We further focus on a model where the inflaton decays to a minimally coupled $\chi$-boson via
the interaction  
\be\label{int}
{\cal L}_{int}=-\fr{1}{2} g^2 \phi^2\chi^2\hs{5}\textrm{or}\hs{5}  {\cal
L}_{int}=-\fr{1}{2} \sigma \phi \chi^2. 
\ee
The first term can give a decay in the broad parametric resonance regime and the
second one is usually considered in the perturbative decay. In either case,  at
least in the first stage until backreaction becomes important the decay process
can be described as the {\it $\chi$-particle creation in a time dependent
classical background}.  One can thus apply the above formulas about particle creation where the time-dependent mass parameter $M$ can be fixed as $M^2=g^2 \phi^2$ or  $M^2=\sigma \phi$. 

In \cite{a3}, we determine the correlation length $\xi_c$ in these two different
reheating models corresponding to the decay in the broad parametric resonance
regime and in the perturbation theory. In the former case, $\xi_c$ is fixed by
the (comoving) momentum scale $k_*$ of the first 
instability band as $\xi_c\sim 1/k_*$. On the other hand, in perturbation theory 
$\xi_c$ is determined up to the scalings due to the expansion of the universe as
$\xi_c\sim 1/m$. In obtaining both results one
simply uses the spectrum of particles produced during  the decay in calculating
the two-point function. Importantly, in each case the correlation length turns
out be smaller than the Hubble scale. 

The above formulas can be used until backreaction effects become important. Prior to backreaction,  linearity is essential and the momentum modes evolve independently without  disturbing locality. However, as the created particles start influencing the background, the linearity is lost and one can no longer treat momentum modes separately. Namely, to determine how the created particles alter the evolution of the background, which necessarily obeys {\it local field equations},  the effects of all the modes must be summed up. Therefore,  it is crucial to determine how physical quantities corresponding to particle creation behave in the real configuration space.  

Consider, for example, the model given by the first interaction term in \eq{int}, which modifies the inflaton field equation as 
 \be\label{ff1}
\ddot{\phi}+3H\dot{\phi}+m^2\phi -g^{ij}(\del_i\del_j\phi)+g^2 \chi^2\phi=0,
\ee
where we also introduce the gradient term for completeness. The backreaction effects become important when the $\chi$-field grows so that  $g^2\chi^2\sim m^2$. In the  Hartree approximation one replaces $\chi^2$ term  in \eq{ff1}  with the expectation value $<\chi^2>$. Since $<\chi^2>$  does not depend on
spatial coordinates and since initially (i.e. just after the end of inflation) the $\phi$-field  is homogenous, only the {\it  zero mode} continues to exist. Thus, in this approximation one can ignore the spatial derivatives in \eq{ff1} and the $\phi$-field remains to be homogenous.   

However, as noted above in \eq{f2} there exists large  fluctuations in $\chi^2$,  comparable to its vacuum expectation value $<\chi^2>$ and thus the "actual value" of $\chi^2$ varies appreciably on scales larger than the correlation length $\xi_c$. To determine the evolution of the inflaton field  in the presence of these fluctuations, one can examine \eq{ff1} in regions of volume $\xi_c^3$, in which $\chi^2$ can be taken as uniform, and try to glue these local results suitably. It is clear that  if only the zero mode survives even when the fluctuations are taken into account, then this should be a good approximation. 

From \eq{ff1}, the frequency of the oscillations  in the $i$'th  region is given $\o_{(i)}^2=m^2+g^2\chi_i^2$,  where $\chi_i^2$ denotes the value of $\chi^2$ in that region and we ignore  the expansion of the universe since in general $m\gg H$ during reheating. This shows that due to fluctuations  in $\chi^2$ given by \eq{f2}  the frequencies also start to differ as  
\be
\fr{\d\o}{\o}\sim \fr{\d\chi^2}{\chi^2}\sim 1.
\ee
In a very short time nearly corresponding to a single average oscillation $t\sim 1/\o\sim1/m$,  the oscillations  of the inflaton field in different regions  become completely  out of phase. In that case, it is no longer permissible to ignore spatial derivatives in \eq{ff1} and the whole reheating dynamics changes completely. 

Actually, the story is more complicated since $\chi^2$ is a random variable. Consider, for instance, the situation when $g^2<\chi^2>$ is about  5\% percent of $m^2$ or so, and thus backreaction effects can be ignored {\it on the average}. However, even at that time one can find regions in which $g^2\chi^2\sim m^2$ and thus coherence of the oscillations is lost near such regions. In other words, one cannot actually talk about a definite time after which backreaction effects become important. 

The above conclusion is specific to the interaction $g^2\phi^2\chi^2$ and one may wonder if it also   holds for other models in general. For example, when the inflaton decays through the interaction $\s\phi\chi^2$, the backreaction term does not modify the frequency of the inflaton oscillations. But this time, there appears a non-homogenous  source term $\s\chi^2$ in the inflaton field equation which alters the amplitude of the oscillations locally. One can see that the fluctuations in $\chi^2$ induce order one changes in the amplitude and again the gradient terms cannot be neglected in the inflaton field equation. 

Thus, we find out that in  reheating models characterized by the interactions \eq{int}, the smoothness of the inflaton background is completely lost and the subsequent evolution must be determined by taking into account the field derivatives. One may think that since the length scale of inhomogeneities is small compared to the cosmologically relevant scales today,  they cannot have any cosmological imprints. However, it is not possible to view these fluctuations as {\it small} perturbations on a homogenous background. Namely, they are expected to alter the evolution of the background in a very non-trivial way (for instance the gradient terms may start dominating  the energy-momentum tensor) and the universe may transform into a state which is completely different than a homogenous and isotropic Robertson-Walker space.  

One may wonder what can be said for other reheating models in general. Until backreaction effects become significant,  \eq{f1}  holds in any model of reheating provided that the decay process can be modeled  as the quantum particle creation in a classical background. Therefore, in such models there inevitably exists order one fluctuations in the energy density and the pressure. As is known, in some  cases the universe can still be described as a perturbed Robertson-Walker metric even though $\d\rho/\rho > {\cal O}(1)$. However, the fluctuations corresponding to the particle creation have interesting properties. For example, it is not possible to characterize them by an equation of state \cite{a3} and consequently  the combination $\r+3 P$, which determine $\ddot{a}$,  would change locally causing different regions to accelerate with different rates.  Moreover, since the size of each region is subhorizon and there exists  large pressure gradients (as shown in \cite{a3}), 
the gravitational collapse can occur exponentially fast as in the case of Jeans instability in Newtonian theory. As a result, one expects these fluctuations to grow and the universe may not enter into a smooth phase in thermal equilibrium. 

In the above discussion we left over some important issues which must actually be studied in realistic scenarios. It is known that for completeness  the metric and the inflaton fluctuations must also be incorporated in the linearized field equations which may alter the particle creation characteristics \cite{mp}. In that case, however, it is not surprising to see the existence of large fluctuations of these fields as well. On the other hand, one may also need to consider the rescattering or the thermalization effects, which may work for uniformization. There are also some conceptual problems one should solve. For example, in \eq{ff1} while $\phi$ is a classical field it is not obvious what  $\chi^2$ should stand for. Here, we take the view that it should be replaced by a  classical realization which would agree with quantum expectation values. However, especially after backreaction sets in, it is not very clear how to determine the evolution of the fields since quantum and classical objects start mixing. All these issues must be resolved before concluding that reheating ends with a perfectly smooth universe in thermal equilibrium. On the contrary, our findings indicate that the opposite is likely to occur.

\end{document}